\documentclass[pra,twocolumn]{revtex4-1}

\usepackage{amsmath,amssymb, bbold, bbm, soul,bm}
\usepackage[pdftex]{graphicx}
\usepackage{epstopdf}
\usepackage{color}
\usepackage{algorithm}

\newcommand{\beq}{\begin{equation}}
\newcommand{\eeq}{\end{equation}}

\newcommand{\ket}[1]{| #1 \rangle}
\newcommand{\bra}[1]{\langle #1 |}

\newcommand{\matele}[3]{\langle #1 | #2 | #3 \rangle}

\newcommand{\by}[1]{#1 \times #1}
\newcommand{\iner}{{\rm In}}

\newcommand{\rk}{{\rm Rank}}
\newcommand{\tr}{{\rm Tr}}
\newcommand{\nn}{\nonumber\\}
\newcommand{\ii}{{\rm i}}
\newcommand{\comment}[1]{}

\begin{document}

\title{Informational completeness in bounded-rank quantum-state tomography}

\author{Charles H. Baldwin}
\email{baldwin4@unm.edu}
\author{Ivan H. Deutsch}
\email{ideutsch@unm.edu}
\author{Amir Kalev}
\email{amirk@unm.edu}
\affiliation{Center for Quantum Information and Control, MSC07--4220, University of New Mexico, Albuquerque, New Mexico 87131-0001, USA}
\date{\today}

\begin{abstract}
We consider the problem of quantum-state tomography under the assumption that the state is pure, and more generally that its rank is bounded by a given value. In this scenario, new notions of informationally complete POVMs emerge, which allow for high-fidelity state estimation with fewer measurement outcomes than are required for an arbitrary rank state. We study this in the context of matrix completion, where the POVM outcomes determine only a few of the density matrix elements. We give an analytic solution that fully characterizes informational completeness and elucidates the important role that the positive-semidefinite property of density matrices plays in tomography. We show how positivity can impose a stricter notion of information completeness and allow us to use convex optimization programs to robustly estimate bounded-rank density matrices in the presence of statistical noise.
\end{abstract}

\maketitle

\section{Introduction} 
Quantum-state tomography (QST) is a standard tool used to characterize, validate, and verify the performance of quantum information processors.  Unfortunately, QST is a demanding experimental task, partly because the number of free parameters of an arbitrary quantum state scale quadratically with the dimension of the system. In most applications, however, the goal is not to manipulate arbitrary states, but to create and coherently evolve pure states. When the device is performing well, and there are only small errors, the quantum state produced will be close to a pure state, and the density matrix will have one a dominant eigenvalue. One can verify the system's performance with other techniques, e.g. randomized benchmarking~\cite{emerson05, knill08, magesan11}, to gain confidence that it is near this regime. This important prior information can be applied to significantly reduce the resources required for QST.  In this work we address this problem by studying QST in the case that the measurement data arises from a density matrix with rank less than or equal to a given value (bounded-rank QST). In particular, we study different aspects of informational completeness that allow for efficient estimation in this scenario, and robust estimation in the face of noise or when the state is full rank but close to a bounded-rank state. 

Bounded-rank QST has been studied by a number of previous workers \cite{Flammia05,Finkelstein04,gross10,liu11,Heinosaari13,Chen13,Goyeneche14,Carmeli14,Kalev15,Carmeli15,Kech15}, and has been shown to require less resources than general QST. One approach is based on the compressed sensing methodology~\cite{gross10,liu11}, where certain sets of randomly chosen measurements guarantee a robust estimation of  low-rank states with high probability. Other schemes~\cite{Flammia05,Finkelstein04,Heinosaari13,Chen13,Goyeneche14,Carmeli15,Kech15}, not related to compressed sensing, construct specific measurements (POVMs) that accomplish bounded-rank QST, and some of these protocols have been implemented experimentally~\cite{Goyeneche14,Hector15}. In addition, some general properties of such measurements have been derived~\cite{Carmeli14, Kalev15, Kech15}.

In bounded-rank QST (rank $\leq r$), two notions of informational completeness become relevant~\cite{Heinosaari13,Chen13,Kalev15}: ``rank-$r$ complete'' and ``rank-$r$ strictly-complete.'' Rank-$r$ complete measurements can distinguish a rank $\leq r$ state from any other rank $\leq r$ state, while rank-$r$ {\em strictly-complete} measurements can distinguish a rank $\leq r$ state from any other physical state, of any rank. Strictly-complete measurements are only possible due to the positive semidefinite property of the density matrix (simply referred to throughout as ``positivity") for bounded-rank QST~\cite{Heinosaari13}. It has been shown~\cite{Kalev15,Kech15-2,Kalev15-2} that strict-completeness has implications for estimating the state of system in the presence of noise. Whereas the set of rank-$r$ states is nonconvex, with rank-$r$ strictly-complete measurements one can use convex optimization programs to estimate the state of the system~\cite{Kalev15,Kech15-2,Kalev15-2}. Under certain conditions~\cite{Kalev15,Kech15-2,Kalev15-2}, such programs allow for robust estimation of the quantum state. Therefore, it is practically advantageous to use strictly-complete measurements for bounded-rank state estimation.

Currently, the literature lacks analytic tools to easily identify and design POVMs that are rank-r strictly-complete, since it is difficult in general to treat the positivity constraint.  We address this problem here by developing new tools based on the concept of the Schur complement. Our method applies to POVMs whose outcomes algebraically determine a particular subset of matrix elements in the density matrix. We refer to such POVMs as ``element-probing'' POVMs (EP-POVMs). For example, the measurements proposed by Flammia~{\em et al.}~\cite{Flammia05} and Goyeneche~{\em et al.}~\cite{Goyeneche14} for pure state tomography are EP-POVMs. In this context, the problem of QST translates to the problem of density matrix completion, where the goal is to recover the entire density matrix when only a few of its elements are given.

The formalism we develop here entirely captures the underlying structure of EP-POVMs and solves the problem of bounded-rank density matrix completion.  As such, it gives methods to test whether a given EP-POVM is a rank-$r$ complete or strictly-complete, and thus it has  applicability to high-fidelity bounded-rank QST.  It also provides intuition on how to design measurement schemes for this task.  Based on this intuition, we construct two examples of rank-$r$ strictly-complete POVMs.  Moreover, our analysis elucidates the role of the positivity constraint in quantum tomography. We will show how to apply the positivity constraint to determine whether an EP-POVM is rank-$r$ strictly complete.

The article is organized as follows. In Sec.~\ref{sec:QST} we review basic concepts in QST and describe in more details the notion of EP-POVMs and the problem of density matrix completion. In Sec.~\ref{sec:formalism} we develop general tools for our method. We show how they can be used to decide whether a given EP-POVM  accomplishes density matrix completion in general and through examples.  In Sec.~\ref{sec:positivity} we discuss the notion of strict-completeness and the role of the positivity constraint in  QST as manifested in this framework. In Sec.~\ref{sec:noise} we discuss the effect of statistical noise on the state reconstruction, and how positivity helps make the estimations robust. In Sec.~\ref{sec:examples} we construct, using the tools developed, two examples of strictly-complete POVMs which can reconstruct states of an arbitrary bounded-rank. These POVMs can be applied iteratively to refine the estimate of the state. Finally, we offer conclusions in Sec.~\ref{sec:conclusions}.

\section{Density matrix completion}\label{sec:QST}
Consider a $d$-dimensional quantum system described by a Hilbert space ${\cal H}_d$, and let ${\cal S}=\{\varrho: \varrho\geq0,\tr(\varrho)=1\}$ be the set of all quantum states on ${\cal H}_d$. A state $\rho$ is a bounded rank-$r$ density matrix if its rank is less than or equal to $r$, i.e., if $\rho\in {\cal S}_r$, where ${\cal S}_r=\{\varrho: \varrho\geq0,\tr(\varrho)=1, \rk(\varrho)\leq r\}$, for a permissible $r$. The set of all quantum states is $ {\cal S}={\cal S}_d$. A quantum measurement on the system is described by a POVM, ${\cal E}=\{E_\mu: E_\mu\geq0, \sum_{\mu=1} E_\mu = \mathbb{1}\}$. The POVM elements,  \{$E_\mu$\}, represent the possible outcomes (events) of the measurement. The probability of measuring an outcome $\mu$ is given by the Born rule $p_\mu = \tr(E_\mu \rho)$. For now, we assume that the system  is measured to infinite precision, i.e., that the probabilities, $\{p_\mu \}$, are known exactly. Noisy measurements will be discussed in Sec~\ref{sec:noise}. 

In general, the task in QST is to solve for the density matrix from the series of linear equations given by the Born rule. When the number of linearly independent POVM elements is equal to $d^2$ (the number of free parameters in a general density matrix) there is a unique solution for the density matrix. In this work we are interested in QST of bounded-rank quantum states, and particularly low-rank quantum states that are described by fewer parameters. We then expect that there exists POVMs with fewer than $d^2$ elements that are informationally complete when the density matrix has bounded rank.

We restrict our attention to EP-POVMs, i.e., POVMs that allow us to algebraically reconstruct density matrix elements. We specifically consider the case where the measurement determines a subset of the total $d^2$ matrix elements, referred to as the ``measured elements.'' With these POVMs, the task of QST is equivalent to the task of density matrix completion, i.e., uniquely reconstructing the remaining (unmeasured) density matrix elements from the measured elements. 

Examples of EP-POVMs were studied by Flammia~{\em et al.}~\cite{Flammia05}, and more recently, by Goyeneche~{\em et al.}~\cite{Goyeneche14} in the context of pure-state tomography. We briefly discuss them here, since we use them as canonical examples for the framework we develop in the next section. Flammia~{\em et al.}~\cite{Flammia05} introduced the following POVM,
\begin{align}\label{psi-complete pure}
&E_0=a\ket{0}\bra{0},\nn
&E_n=b(\mathbb{1}+\ket{0}\bra{n}+\ket{n}\bra{0}),\; \;n=1,\ldots,d-1,\nn
&\widetilde{E}_n=b(\mathbb{1}-\ii\ket{0}\bra{n}+\ii\ket{n}\bra{0}),\; \;n=1,\ldots,d-1,\nn
&E_{2d}=\mathbb{1}-\Bigl[E_0 +\sum_{n=1}^{d-1}(E_n+\widetilde{E}_n)\Bigr],
\end{align}
with $a$ and $b$ chosen such that $E_{2d}\geq0$.  They showed that the measurement probabilities $p_\mu=\bra{\psi}E_\mu\ket{\psi}$ and $\tilde{p}_\mu=\bra{\psi}\widetilde{E}_\mu\ket{\psi}$ can be used to reconstruct any $d$-dimensional pure state $\ket{\psi}=\sum_{k=0}^{d-1}c_k\ket{k}$, as long as  $c_0\neq0$. Under the assumption, $c_0>0$, we find that  $c_0=\sqrt{p_0}/a$. The real and imaginary parts of $c_n$, $n=1,\ldots, d-1$, are found through the relations $\Re(c_n)=\frac1{2c_0}(\frac{p_n}{b}-1)$ and $\Im(c_n)=\frac1{2c_0}(\frac{\tilde{p}_n}{b}-1)$, respectively. The POVM in Eq.~\eqref{psi-complete pure} is in fact an EP-POVM where the measured elements are the first row and column of the density matrix. The probability $p_0=\tr(E_0\rho)$ can be used to algebraically reconstruct the element $\rho_{0,0}=\matele{0}{\rho}{0}$, and the probabilities $p_n=\tr(E_n\rho)$ and $\tilde{p}_n=\tr(\widetilde{E}_n\rho)$ can be used to reconstruct the elements $\rho_{n,0}=\matele{n}{\rho}{0}$ and $\rho_{0,n}=\matele{0}{\rho}{n}$, respectively. 

A different POVM for pure state tomography was studied by Goyeneche~{\em et al.}~\cite{Goyeneche14}. In this scheme four specific orthogonal bases are measured, 
\begin{align}\label{4gmb}
\mathbbm{B}_{1} &=\Bigl\{ \frac{\ket{0}\pm\ket{1}}{\sqrt2}, \frac{\ket{2}\pm\ket{3}}{\sqrt2}, \ldots, \frac{\ket{d-2}\pm\ket{d-1}}{\sqrt2}\Bigr\}, \nonumber \\
\mathbbm{B}_{2} &=\Bigl\{ \frac{\ket{1}\pm\ket{2}}{\sqrt2}, \frac{\ket{3}\pm\ket{4}}{\sqrt2}, \ldots, \frac{\ket{d-1}\pm\ket{0}}{\sqrt2}\Bigr\}, \nonumber \\
\mathbbm{B}_{3} &=\Bigl\{ \frac{\ket{0}\pm\ii\ket{1}}{\sqrt2}, \frac{\ket{2}\pm\ii\ket{3}}{\sqrt2}, \ldots, \frac{\ket{d-2}\pm\ii\ket{d-1}}{\sqrt2}\Bigr\}, \nonumber \\
\mathbbm{B}_{4} &=\Bigl\{ \frac{\ket{1}\pm\ii\ket{2}}{\sqrt2}, \frac{\ket{3}\pm\ii\ket{4}}{\sqrt2}, \ldots, \frac{\ket{d-1}\pm\ii\ket{0}}{\sqrt2}\Bigr\}.
\end{align}
Goyeneche~{\em et al.}~\cite{Goyeneche14} outlined a procedure to reconstruct the pure state amplitudes but we omit it here for brevity. Similar to the POVM in Eq.~\eqref{psi-complete pure}, the procedure fails when certain state-vector amplitudes vanish. More details are given in Sec.~\ref{sec:examples}. This POVM is an EP-POVM as well. Here, the measured elements are the elements on the first diagonals (the diagonals above and below the principal diagonal) of the density matrix. Denoting $p_{j}^{\pm}=\frac{1}{2}(\bra{j}\pm\bra{j+1})\rho(\ket{j}\pm\ket{j+1})$, and $p_{j}^{\pm\ii}=\frac{1}{2}(\bra{j}\mp\ii\bra{j+1})\rho(\ket{j}\pm\ii\ket{j+1})$, we obtain, $\rho_{j,j+1}{=}\frac{1}{2}[(p_{j}^{+}-p_{j}^{-})+\ii (p_{j}^{+\ii}-p_{j}^{-\ii})]$ for $j = 0,\ldots,d-1$, and addition of indices is taken modulo $d$.

Goyeneche~{\em et al.}~\cite{Goyeneche14} also considered a protocol for pure-state tomography by adaptively measuring five bases. In Sec~\ref{ssec:examples_bases} we consider a related protocol with five-bases but without adaptation.

By their design, these two POVMs can distinguish a pure state form any other pure state by a well-chosen construction. However, currently there is no unified and simple description of the underlying structure of POVMs that allow for a pure-state, and more generally bounded-rank state, identification. Moreover, due to the positivity constraint it is generally difficult to determine if there are other density matrices of higher rank that are consistent with the POVM probabilities. We address these issues in the subsequent sections by developing a framework that accomplishes bounded-rank QST in the context of EP-POVMs and deals with the positivity constraint explicitly. We note that while our discussion is presented in terms of quantum states, the framework is broader. The trace constraint on the matrix plays no role in the framework below, and thus applies to any positive semidefinite matrices, not solely those with unit trace (i.e., quantum states).

\section{Rank-$r$ complete POVMs}\label{sec:formalism}
\begin{figure}
\centering
\includegraphics[width=\linewidth]{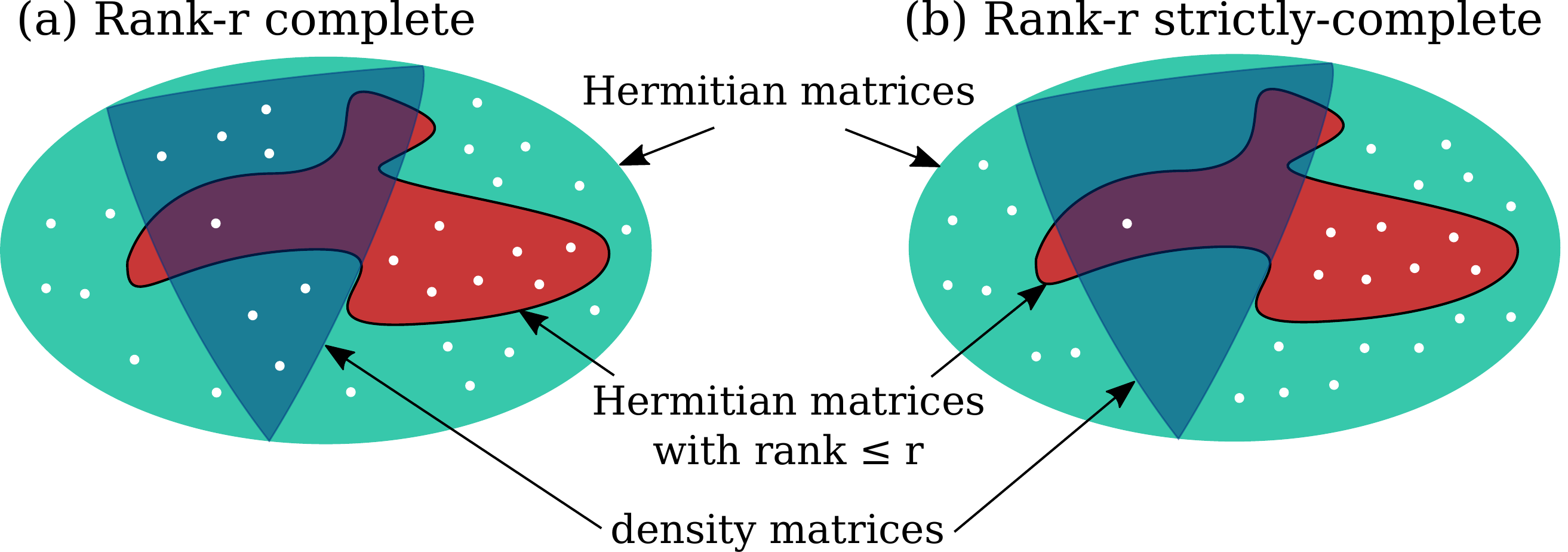}
\caption{{\bf Two notions of informational completeness for measurements of a rank-$r$ quantum state.} Three sets are denoted schematically: The set of all $d\times d $ Hermitian  matrices (green);  the nonconvex set of Hermitian matrices with rank $\leq r$ (red); the convex subset of positive trace-1 matrices, i.e., all physical density matrices (blue).  The white points indicate Hermitian matrices which are consistent with the (noiseless) measurement record. {\bf(a) Rank-$r$ complete POVM.} The measurement  uniquely specifies a rank-$r$ density matrix from within the set of rank-$r$ density matrices. It does not uniquely indentify it from other density matrices with rank greater than $r$. {\bf (b) Rank-$r$ strictly-complete POVM.}  The measurement  uniquely specifies a rank-$r$ density matrix from within the set of all density matrices. The existence of such nontrivial POVMs is due to the positivity property of the density matrix. The only higher-rank matrices consistent with the measurement record are strictly negative, and thus not physical.}
\label{fig:illustration}
\end{figure}

A formal definition of informational completeness which allows for unique recovery of a bounded-rank quantum states is given as follows~\cite{Heinosaari13,Kalev15}.\\
\\
{\noindent{\bf Definition~1.} A POVM is said to be {\em rank-$r$ complete} if no two distinct states $\rho$ and $\sigma$ in ${\cal S}_r$ yield the same measurement probabilities,
\begin{equation} 
\label{restricted_definition}
\forall \, \rho, \sigma \in {\cal S}_r, \rho\neq \sigma, \, \exists \,  E_\mu\in{\mathcal E} \,\, {\rm {\rm s.t.}} \,\, \tr(E_\mu \rho) \neq \tr(E_\mu \sigma),
\end{equation}
except for a set of rank-$r$ states that are dense on a set of measure zero, called the ``failure set.''
}
\\\\
By definition, the probabilities of a rank-$r$ complete POVM uniquely identify a rank-$r$ state from within the set of rank-$r$ states. Thus, if an EP-POVM is rank-$r$ complete, then the measurement probabilities uniquely distinguish the state from any other rank-$r$ state~\cite{footnote_tr}, see Fig.~\ref{fig:illustration}(a). For $r = 1$, i.e., pure states, Definition~1 coincides with the definition of pure-state informationally-complete (PSI-complete) of Ref.~\cite{Flammia05}. For example, the POVMs proposed by Flammia~{\em et al.} and by Goyeneche~{\em et al.} are used to algebraically reconstruct the amplitudes of pure states, and therefore they are rank-1 complete POVMs. We comment on the implications of the failure set where appropriate.

\subsection{Density matrix completion and the Schur complement}\label{ssec:technique}
Our technique to determine whether an EP-POVM accomplishes density matrix completion relies on properties of the Schur complement~\cite{Haynsworth68,Zhang11}. Consider a block-partitioned $\by{k}$ Hermitian matrix $M$,
\begin{equation} \label{block_mat}
M = 
\begin{pmatrix}
{A} & {B^{\dagger}} \\
{B} &{C} 
\end{pmatrix},
\end{equation}
where $A$ is a $\by{r}$ Hermitian matrix, and the size of ${B^{\dagger}}$, $B$ and $C$ is determined accordingly.
The Schur complement of $M$ with respect to $A$, assuming $A$ is nonsingular, is defined by
\begin{equation}
M/A \equiv C - B A^{-1} B^{\dagger}.
\end{equation}
We will use the Haynsworth inertia (In) additivity formula, which relates the inertia of $M$ to that of $A$ and of $M/A$~\cite{Haynsworth68},
\begin{equation} \label{Schur_iner}
\iner(M) = \iner(A)+\iner(M/A),
\end{equation}
where the inertia of a Hermitian matrix, $H$, is the ordered triple of the number of negative, zero, and positive eigenvalues of $H$, ${\rm In}(H)=(n_-, n_0, n_+)$, respectively. A corollary of the inertia formula is the rank additivity property,
\begin{equation} \label{Schur_rank}
\rk(M) = \rk(A) + \rk(M/A).
\end{equation}
With these relations we can determine if any EP-POVM is rank-$r$ complete.

As an instructive example, we use the properties of the Schur complement, in an alternative proof that the POVM in Eq.~\eqref{psi-complete pure} is rank-1 complete, without referring to the pure-state amplitudes. The POVM in Eq.~\eqref{psi-complete pure} is an EP-POVM, where the measured elements are $\rho_{0,0}$, $\rho_{n,0}$ and $\rho_{0,n}$ for $n=1,\ldots,d-1$. Supposing that $\rho_{0,0}>0$ and labeling the unmeasured $\by{(d-1)}$ block of the density matrix by $C$, we write
\begin{equation} \label{block_rho}
\rho=  \left(
    \begin{array}{cccc}
{\rho_{0,0}} &  {\rho_{0,1}}& \cdots &{\rho_{0,d-1}}\\
\cline{2-4}\multicolumn{1}{c|}{\rho_{1,0}}&
      {} &{}& \multicolumn{1}{c|}{} \\
      \multicolumn{1}{c|}{\vdots}&
      {} &{\;\;\Large\textit{C}}& \multicolumn{1}{c|}{}\\
\multicolumn{1}{c|}{\rho_{d-1,0}}&
      {} &{}& \multicolumn{1}{c|}{}\\\cline{2-4}
    \end{array}
    \right)
\end{equation}
Clearly, Eq.~\eqref{block_rho} has the same form as Eq.~\eqref{block_mat}, such that $M = \rho$, $A = \rho_{0,0}$, $B^{\dagger}=({\rho_{0,1}}\cdots {\rho_{0,d-1}})$, and $B = ({\rho_{0,1}}\cdots {\rho_{0,d-1}})^{\dagger}$. Assume $\rho$ is a pure state so $\rk(\rho)=1$. By applying Eq.~\eqref{Schur_rank} and noting that $\rk(A)=1$, we obtain $\rk(\rho/A)=0$. This implies that $\rho/A=C - B A^{-1} B^{\dagger}=0$, or equivalently, that $C= B A^{-1} B^{\dagger}= \rho_{0,0}^{-1}B B^{\dagger}$. Therefore, by measuring every element of $A$, $B$ (and thus of ${B^{\dagger}}$), the rank additivity property allows us to algebraically reconstruct $C$ uniquely without measuring it directly. Thus, the entire density matrix is determined by measuring its first row and column.

Here, we used the assumption that $\rk(\rho){=}1$. Therefore, based on this analysis, the reconstructed state is unique to the set ${\cal S}_1$, and the POVM is rank-1 complete. There might, however, be other density matrices of rank greater than one that have the same measured blocks $A$, ${B^{\dagger}}$ and $B$ but are different on the unmeasured block $C$. Moreover, this algebraic reconstruction of the rank-$1$ density matrix works as long as $\rho_{0,0}\neq0$. When $\rho_{0,0}=0$, the Schur complement is not defined, and Eq.~\eqref{Schur_rank} does not apply. This, however, only happens on a set of states of measure zero (the failure set), i.e. the set of states where $\rho_{0,0} = 0$ exactly. It is the exact same set found by Flammia~{\em et al.}~\cite{Flammia05}.

The above technique can be used to determine if any EP-POVM is rank-$r$ complete for a state $\rho\in{\cal S}_r$. In general, the structure of the measured elements will not be as convenient as the example considered above. Our approach is to study $\by{k}$ principle submatrices of $\rho$ such that $k > r$. Since $\rho$ is a rank-$r$ matrix, it has at least one nonsingular $\by{r}$ principal submatrix~\cite{footnote_rank}. Assume, for now, that a given $\by{k}$ principal submatrix, $M$, contains a nonsingular $\by{r}$ principle submatrix $A$. From Eq.~\eqref{Schur_rank}, since $\rk(M) = \rk(A) = r$, $\rk(M/A) = 0$, and therefore, $C = B A^{-1}B^{\dagger}$. This equation motivates our choice of $M$. If the measured elements make up $A$ and $B$ (and therefore $B^{\dagger}$) then we can solve for $C$ and we have fully characterized $\rho$ on the subspace defined by $M$. We refer to block-matrices in this form as a principal submatrix in the canonical form. In practice, the measured elements only need to be related to canonical form by a unitary transformation. In Sec.~\ref{ssec:examples_bases} we discuss such an example where the transformation is done by interchanging columns and corresponding rows. In general, an EP-POVM may measure multiple subspaces, $M_i$, and we can reconstruct $\rho$ only when the corresponding $A_i$, $B_i$, $C_i$ cover all elements of $\rho$~\cite{footnote_tr}. We label the set of all principle submatrices that are used to construct $\rho$ by ${\cal M} = \{M_i\}$. Since we can reconstruct a unique state within the set of $\mathcal{S}_r$ this is then a general description of a rank-$r$ complete EP-POVM.  The failure set, in which the measurement fails to reconstruct $\rho$, corresponds to the set of states that are singular on any of the $A_i$ subspaces. 

\section{Rank-$r$ strictly-complete POVMs: The role of positivity}\label{sec:positivity}
The definition of rank-$r$ complete POVM guarantees the uniqueness of the reconstructed state in the set ${\cal S}_r$, but it does not say anything about higher-rank states. There may be other density matrices, with rank greater than $r$ that are consistent with the measurement record (or with the measured elements). Since ${\cal S}_r$ is a nonconvex set, in practice it may be difficult to differentiate between the unique rank-$r$ density matrix and these higher-rank states. We can, however, consider a ``stricter" type of POVMs which excludes these higher-rank states.  This motivates the following definition~\cite{Chen13,Carmeli14,Kalev15}.\\
\\
{\noindent{\bf Definition~2.}  A POVM is said to be {\em rank-$r$ strictly-complete} if no two distinct states $\rho\in{\cal S}_r$ and $\sigma\in{\cal S}$ yield the same measurement probabilities,
\begin{equation} 
\label{strictly_definition}
\forall \rho\in{\cal S}_r, \forall\sigma \in\mathcal{S}, \rho\neq \sigma, \, \exists \, E_\mu\in{\mathcal E} \, {\rm {\rm s.t.}} \, \tr(E_\mu \rho) \neq \tr(E_\mu \sigma),
\end{equation}
except for a set of rank-$r$ states that are dense on a set of measure zero, called the ``failure set.''
}
\\\\
Clearly, if a POVM is rank-$r$ strictly-complete, it is also rank-$r$ complete. If the state has rank less than or equal to $r$, then a rank-$r$ strictly-complete POVM uniquely identifies it from within the set all quantum states, see Fig.~\ref{fig:illustration}(b). This makes rank-$r$ strictly-complete POVMs ideal for use with convex optimization, more details are given in Sec.~\ref{sec:noise}. If an EP-POVM is rank-$r$ strictly-complete, then the measurement probabilities uniquely determine the rank-$r$ state from within the set of all quantum states. There are no higher-rank density matrices that have the same measurement probabilities~\cite{footnote_tr}.

The notion of strict-completeness is only nontrivial due to the positivity constraint of quantum states. To see this, let us ignore the positivity constraint, only constraining the density matrix to be Hermitian, and apply the definition of strict-completeness for bounded-rank Hermitian matrices. Let $R$ be a Hermitian matrix with ${\rm Rank}(R)\leq r$. To be (nontrivially) strictly-complete the POVM should be able to distinguish $R$ from any Hermitian matrix, of any rank, with less than $d^2$ linearly independent POVM elements. (If the POVM has $d^2$ linearly independent POVM elements, it is fully informationally complete and can distinguish any Hermitian matrix from any other.) However, a POVM with less than $d^2$ linearly independent elements necessarily has infinitely many Hermitian matrices, with rank $> r$, which produce the same noiseless measurement record as $R$. Therefore, without positivity, we cannot define strict-completeness with less than $d^2$ linearly independent elements. On the other hand, if we impose positivity, as we will shortly see, there exists POVMs that are rank-$r$ strictly-complete and require fewer than $d^2$ elements.

As an example, we show that under the positivity requirement the POVM in Eq.~\eqref{psi-complete pure} is in fact a rank-$1$ strictly-complete POVM. We already know that the POVM in Eq.~\eqref{psi-complete pure} is rank-$1$ complete and thus $\rho/A=0$. By applying the inertia additivity formula to $\rho$ we obtain 
\begin{equation}
\iner(\rho) = \iner(A)+\iner(\rho/A)=\iner(A).
\end{equation}
This implies that $A$ is a positive semidefinite matrix. For the POVM in Eq.~\eqref{psi-complete pure} $A=\rho_{0,0}$, so this equation is a re-derivation of the trivial condition $\rho_{0,0}\geq0$. Let us assume that the POVM is not rank-$1$ strictly-complete. If so, there must exist a quantum state, $\sigma\geq0$, with $\rk(\sigma)>1$, that has the same measurement record and thus measured elements as $\rho$, but different unmeasured elements. We define this difference by $V\neq0$, and write
\begin{equation} \label{block_mat_sigma}
\sigma=  \left(
    \begin{array}{cccc}
{\rho_{0,0}} &  {\rho_{0,1}}& \cdots &{\rho_{0,d-1}}\\
\cline{2-4}\multicolumn{1}{c|}{\rho_{1,0}}&
      {} &{}& \multicolumn{1}{c|}{} \\
      \multicolumn{1}{c|}{\vdots}&
      {} &{\;\;\Large{\textit{C}}+\!\Large{\textit{V}}}& \multicolumn{1}{c|}{}\\
\multicolumn{1}{c|}{\rho_{d-1,0}}&
      {} &{}& \multicolumn{1}{c|}{}\\\cline{2-4}
    \end{array}
    \right)=\rho+\begin{pmatrix}
{0} &  {\bf 0}\\
{\bf 0} & V
\end{pmatrix}.
\end{equation}
Since $\sigma$ and $\rho$ have the same measurement record, for all $\mu$, $\tr(E_\mu\sigma)=\tr(E_\mu\rho)$. Summing over $\mu$ and using $\sum_\mu E_\mu=\mathbb{1}$, we obtain that $\tr(\sigma)=\tr(\rho)$. This implies that $V$ must be a traceless Hermitian matrix, hence, $n_-(V) \geq 1$. Using the inertia additivity formula for $\sigma$ gives,
\begin{equation}
\iner(\sigma) = \iner(A)+\iner(\sigma/A).
\end{equation}
By definition, the Schur complement
\begin{equation}
\sigma/A=C +V - B A^{-1} B^{\dagger}=\rho/A+V=V.
\end{equation}
The inertia additivity formula for  $\sigma$ thus reads,
\begin{equation}
\iner(\sigma) = \iner(A)+\iner(V).
\end{equation}
Since $A=\rho_{0,0}>0$, $n_-(\sigma) = n_-(V) \geq 1$ so $\sigma$ has at least one negative eigenvalue, in contradiction to the assumption that it is a positive semidefinite matrix. Therefore, $\sigma \ngeq 0$ and we conclude that the POVM in Eq.~\eqref{psi-complete pure} is rank-1 strictly-complete. 

A given POVM that is rank-$r$ complete is not necessarily rank-$r$ strictly-complete like the POVM in Eq.~\eqref{psi-complete pure}. For example, the bases in Eq.~\eqref{4gmb}, correspond to a rank-$1$ complete POVM, but not to a rank-$1$ strictly-complete POVM. For these bases, we can apply a similar analysis to show that there exists a quantum state $\sigma$  with $\rk(\sigma)>1$ that matches the measured elements of $\rho$.

We now derive the necessary and sufficient condition for a rank-$r$ complete EP-POVMs to be rank-$r$ strictly-complete. Using the notation introduced in Sec.~\ref{ssec:technique}, let us choose an arbitrary principal submatrix $M\in {\cal M}$ that was used to construct $\rho$. Such a matrix has the form of Eq.~\eqref{block_mat} where $C=BA^{-1}B^\dagger$. Let $\sigma$ be a higher-rank matrix that has the same measured elements as $\rho$, and let $\tilde{M}$ be the submatrix of $\sigma$ which spans the same subspace as $M$. Since $\sigma$ has the same measured elements as $\rho$, $\tilde{M}$ must have the form
\begin{equation}\label{Mtilde}
\tilde{M} = 
\begin{pmatrix}
 A & B^{\dagger} \\
 B & \tilde{C} 
 \end{pmatrix}\equiv\begin{pmatrix}
 A & B^{\dagger} \\
 B & C + V 
 \end{pmatrix}=M+\begin{pmatrix}
 {\bf 0} & {\bf 0} \\
 {\bf 0} & V 
 \end{pmatrix}.
 \end{equation}
Then, from Eq.~\eqref{Schur_iner}, $\iner(\tilde{M}) = \iner(A) + \iner(\tilde{M}/A) = \iner(A) + \iner(V)$, since $\tilde{M}/A = M/A + V = V$. If $\sigma$ is a density matrix then it is a positive semidefinte matrix. A matrix is positive semidefinite if and only if all of its principal submatrices are positive semidefinite~\cite{Zhang11}. Therefore, $\sigma \geq 0$ if and only if $\tilde{M}\geq 0$, and $\tilde{M} \geq 0$ if and only if $n_-(A) + n_-(V) = 0$. Since $\rho\geq0$, all its principal submatrices are positive semidefinite, and in particular $A\geq0$. Therefore,  $\sigma$ is a state if and only if $n_-(V) = 0$. We can repeat this logic for all other submatrices $M \in \mathcal{M}$. Hence we conclude that the measurement is rank-$r$ strictly-complete if and only if there exists at least one submatrix $M \in \mathcal{M}$ for which every $V$ that we may add (similarly to Eq.~\eqref{Mtilde}) has at least one negative eigenvalue. 

A sufficient condition for an EP-POVM to be rank-$r$ strictly-complete is given in the following proposition.\\
\\
{\noindent{\bf Proposition~1.} Assume that an EP-POVM is rank-$r$ complete. If it measures the diagonal elements of the density matrix, then it is rank-$r$ strictly-complete.}\\
{\em Proof.} Consider a Hermitian matrix $\sigma$ that has the same measurement probabilities as $\rho$, thus the same measured elements.  If we measure all diagonal elements of $\rho$ (and thus, of $\sigma$), then for any principal submatrix $\tilde{M}$ of $\sigma$, {\em cf.} Eq.~\eqref{Mtilde}, the corresponding $V$ is traceless because all the diagonal elements of $C$ are measured. Since $V$ is Hermitian and traceless it must have at least one negative eigenvalue, therefore, $\sigma$ is not positive semidefinite matrix and the POVM is rank-$r$ strictly-complete.\hfill $\square$ \\
\\
A useful corollary of this proposition is any EP-POVM that is rank-$r$ complete can be made rank-$r$ strictly-complete simply by measuring the diagonal elements of the density matrix.

\section{Tomography in the presence of noise}\label{sec:noise}
So far, we have discussed the ideal noiseless case but in this section we will consider the affects of realistic noise on bounded-rank QST. We discuss two important sources: statistical noise in the measurements and preparation noise. Statistical noise can come from finite sampling of the quantum system or other random fluctuations. Preparation can come from inhomogenities in control or decoherence and causes the state being measured to not be represented by a bounded-rank density matrix.

Either source of noise may cause the algebraic reconstruction presented above to not return a density matrix, i.e, a positive semidefinite matrix. In this case, we typically use numerical programs to obtain an estimation by minimizing a cost function under appropriate constraints. The cost function, for example, can be a distance measure between the measurement record and the model. The result would be a density matrix that estimates the measurement record, according to the cost function that we minimized. If the POVM we use is rank-$r$ complete, there are potentially other higher-rank density matrices which produce similar data, but are very different than the (bounded-rank) state of the system. Therefore, we must search for an estimate only within the set of rank-$r$ states. However, the set of rank-$r$ states is a nonconvex set and, therefore, it is difficult to find global optimum. Hence, in practice it is difficult to reliably estimate bounded-rank quantum states by measuring rank-$r$ complete POVMs. 

On the contrary, rank-$r$ strictly-complete POVMs uniquely identify, in the absence of noise, a rank-$r$ state from within the set of {\em all} density matrices. Since the set of density matrices is a convex set we can use convex programs to estimate the state. Convex programs are appealing since they are efficient and converge to global optimum. Moreover, it was recently shown~\cite{Kalev15,Kech15-2,Kalev15-2} that the estimate returned by a convex program is robust to both sources of noise. The positivity constraint allows us to form rank-$r$ strictly-complete POVMs, which in turn, enables us to treat the problem of bounded-rank QST as a convex problem and thereby estimate the state in a robust way.

Another implication of statistical noise is on the failure set of a given POVM. As described above, this set correspond to states for which a principal submatrix $A_i$ used in the reconstruction is singular. Without noise, this is a set of measure zero and is avoided with probability one. However, in the presence of statistical noise this is no longer true. Consider the POVM in Eq.~\eqref{psi-complete pure} in the absence of statistical noise, its failure set corresponds to all states for which $\rho_{00}=0$. If the statistical noise has fluctuations of the order of $\epsilon$ then every state within a finite ball $\rho_{00}\lesssim\epsilon$ cannot be distinguished from a state with $\rho_{00}=0$, and thus could not be identified by this POVM~\cite{Finkelstein04}. While the POVM in Eq.~\eqref{psi-complete pure} has a failure set with a simple structure, other POVMs may have failure sets that are more complex. The more complicated the structure, the less likely it is to affect the estimation of the state in the presence of statistical noise. We consider such an example in Sec.~\ref{ssec:examples_bases}. We thus do not regard the failure set as a practical limitation.

\section{Construction of rank-$r$ strictly-complete POVMs}\label{sec:examples}
The framework we developed in the previous sections allows us to construct rank-$r$ strictly-complete POVMs. We present two such POVMs, and describe the algebraic reconstruction of the rank-$r$ state. The POVMs are generalization of the POVMs by Flammia~{\em et al.}~\cite{Flammia05} and Goyeneche~{\em et al.}~\cite{Goyeneche14} from pure states to rank-$r$ states. The construction of these POVMs is made easier thanks to the tools presented in this work. Additionally, these POVMs can be implemented iteratively, since the construction of rank-$(r-1)$ strictly-complete POVM is nested in rank-$r$ strictly-complete POVM. Therefore, suppose that it is {\em a priori} known that the state is nearly pure, then one could estimate it based on a rank-$1$ strictly-complete POVM. If the estimate is not satisfactory, or if we wish to learn about the state beyond the rank-1 estimation, one can, e.g., complement the rank-$1$ strictly-complete POVM, into a rank-2 strictly-complete POVM by a few additional measurements.

\subsection{Example~1}\label{ssec:examples_full}
A rank-$r$ density matrix has $(2d-r)r-1$ free parameters. The first rank-$r$ strictly-complete POVM we form has $(2d-r)r+1$ elements, and is a generalization of ther POVM in Eq.~\eqref{psi-complete pure}. The POVM elements are,
\begin{align}\label{psic mixed}
&E_k=a_k\ket{k}\bra{k},\;k=0,\ldots,r-1\nn
&E_{k,n}=b_k(\mathbb{1}+\ket{k}\bra{n}+\ket{n}\bra{k}),\;n=k+1,\ldots,d-1,\nn
&\widetilde{E}_{k,n}=b_k(\mathbb{1}-\ii\ket{k}\bra{n}+\ii\ket{n}\bra{k}), \;n=k+1,\ldots,d-1,\nn
&E_{(2d-r)r+1}=\mathbb{1}-\sum_{k=0}^{r}\Bigl[E_k +\sum_{n=1}^{d-1}(E_{k,n}+\widetilde{E}_{k,n})\Bigr],
\end{align}
with $a_k$ and $b_k$ chosen such that $E_{(2d-r)r+1}\geq0$. The probability $p_k=\tr(E_k\rho)$ can be used to calculate the density matrix element $\rho_{k,k}=\matele{k}{\rho}{k}$, and the probabilities $p_{k,n}=\tr(E_{k,n}\rho)$ and $\tilde{p}_{k,n}=\tr(\widetilde{E}_{k,n}\rho)$ can be used to calculate the density matrix elements $\rho_{n,k}=\matele{n}{\rho}{k}$ and $\rho_{k,n}=\matele{k}{\rho}{n}$. Thus, this is an EP-POVM which reconstruct the first $r$ rows and first $r$ columns of the density matrix. 

Given the measured elements, we can write the density matrix in block form corresponding to measured and unmeasured elements,
\begin{equation} \label{block_rho_gen}
\rho= 
\begin{pmatrix}
A &  B^\dagger\\
B & C 
\end{pmatrix},
\end{equation}
where $A$ is a $\by{r}$ submatrix and $A$, $B^\dagger$, and $B$ are composed of measured elements. Suppose that $A$ is nonsingular. Given that $\rk(\rho)=r$, using the rank additivity property of Schur complement and that $\rk(A)=r$, we obtain $\rho/A=C-BA^{-1}B^\dagger=0$. Therefore, we conclude that $C=BA^{-1}B^\dagger$.  Thus we can reconstruct the entire rank-$r$ density matrix. 

Following the arguments for the POVM in Eq.~\eqref{psi-complete pure}, it is straight forward to show that this POVM is in fact rank-$r$ strictly-complete. The failure set of this POVM corresponds to states for which $A$ is singular. The set is dense on a set of states of measure zero. 

The POVM of Eq.~\eqref{psic mixed} can alternatively be implemented as a series of $r-1$ POVMs, where the $k$th POVM, $k=0,\ldots,r-1$, has $2(d-k)$ elements, 
\begin{align}\label{psic mixed kth}
&E_k=a_k\ket{k}\bra{k},\nn
&E_{k,n}=b_k(\mathbb{1}+\ket{k}\bra{n}+\ket{n}\bra{k}),\;n=k+1,\ldots,d-1,\nn
&\widetilde{E}_{k,n}=b_k(\mathbb{1}-\ii\ket{k}\bra{n}+\ii\ket{n}\bra{k}), \;n=k+1,\ldots,d-1,\nn
&E_{2(d-k)}=\mathbb{1}-\Bigl[E_k +\sum_{n=1}^{d-1}(E_{k,n}+\widetilde{E}_{k,n})\Bigr].
\end{align}
Suppose that the state of the system is close to a pure state. By measuring the $k=0$ POVM,  which is the  POVM in Eq.~\eqref{psi-complete pure}, we obtain most  of  the  information  about  the state.  If we continue and measure the $k=1$ POVM, we will obtain the ``first-order'' correction for the estimate. In the same way, we can measure higher $k$ POVMs and obtain ``higher-order'' estimates of the state.

\subsection{Example~2}\label{ssec:examples_bases}
The second rank-$r$ strictly-complete POVM we assemble corresponds to a measurement of $4r+1$ orthonormal  bases, which is a generalization of the four basis in Eq.~\eqref{4gmb}. We consider the case that the dimension of the system is a power of two. Since a measurement of $d+1$ mutually unbiased bases is informationally complete~\cite{Wootters89}  (can distinguish any quantum state from any other), this construction is relevant as long as $r< d/4$.  We first asses the case of $r=1$, which is similar to the measurement proposed by Goyeneche~{\em et al.}~\cite{Goyeneche14}. In this case there are five bases, the first is the computational basis, $\{\ket{k}\}$, $k=0,\ldots,d-1$ and the other four are given in Eq.~\eqref{4gmb}. Goyeneche~{\em et al.}~\cite{Goyeneche14} showed that the last four bases are rank-1 complete. Here, we show these five bases are rank-1 strictly-complete with the techniques introduced above.

We label the upper-right diagonals $0$ to $d-1$, where the $0$th diagonal is the principal diagonal and the $(d-1)$st diagonal is the upper right element. Each diagonal, except the $0$th, has a corresponding Hermitian conjugate diagonal (its corresponding lower-left diagonal). Thus, if we measure the elements on a diagonal, we also measure the elements of its Hermitian conjugate. The computational basis corresponds to measuring the $0$th diagonal. In Sec.~\ref{sec:QST} we showed measuring the last four bases corresponds to measuring the elements on the first  diagonals. To show that the measurement of these five bases is rank-1 complete, we follow a similar strategy outlined in Sec.~\ref{sec:formalism}. First, choose the leading $\by{3}$ principal submatrix,
\begin{equation}
M_0 = 
\begin{pmatrix}
\rho_{0,0} & \rho_{0,1} & \bm{\rho_{0,2}} \\
\rho_{1,0} & \rho_{1,1} & \rho_{1,2} \\
\bm{\rho_{2,0}} & \rho_{2,1} &\rho_{2,2}  \\
\end{pmatrix},
\end{equation}
where, hereafter, the elements in bold font are the unmeasured elements. By applying a unitary transformation, which switches the first two rows and columns, we can move $M_0$ into the canonical form, 
\begin{equation}
M_0 \rightarrow UM_0U^\dagger=
\begin{pmatrix}
\rho_{1,1} &\rho_{1,0} & \rho_{1,2} \\
\rho_{0,1} & \rho_{0,0} & \bm{\rho_{0,2}} \\
 \rho_{2,1} & \bm{\rho_{2,0}} & \rho_{2,2}   \\
\end{pmatrix}.
\end{equation}
From Eq.~\eqref{Schur_rank} we can solve for the bottom $\by{2}$ block of $UM_0U^\dagger$ if $\rho_{1,1} \neq 0$. The set of states with $\rho_{1,1}= 0$ corresponds to the failure set. Note that the diagonal elements of the bottom $\by{2}$ block, $\rho_{0,0}$ and $\rho_{2,2}$, are also measured. We repeat this procedure for the set of principal $\by{3}$ submatrices, $M_{i} \in \mathcal{M}$, $i=0,\ldots,d-2$,
\begin{equation}
M_{i} = \begin{pmatrix}
\rho_{i,i} & \rho_{i,i+1} & \bm{\rho_{i,i+2}} \\
\rho_{i+1,i} & \rho_{i+1,i+1} & \rho_{i+1,i+2} \\
\bm{\rho_{i+2,i}} & \rho_{i+2,i+1} &\rho_{i+2,i+2}  \\
\end{pmatrix},
\end{equation}
For each $M_{i}$, the upper-right and the lower-left corners elements $\rho_{i,i+2}$ and $\rho_{i+2,i}$ are unmeasured. Using the same procedure as above we reconstruct these elements for all values of $i$ and thereby reconstruct the 2nd diagonals. We repeat the entire procedure again choosing a similar set of $\by{4}$ principal submatrices and reconstruct the 3rd diagonals and so on for the rest of the diagonals until all the unknown elements of the density matrix are reconstructed. Since, we  have reconstructed all diagonal elements of the density matrix and used the assumption that $\rk{(\rho)} = 1$ these five bases correspond to rank-$1$ complete POVM. The first basis measures the 0th diagonal so by Proposition~1 the measurement is rank-1 strictly-complete.

The failure set corresponding to ${\cal M}$ is when $\rho_{i,i} = 0$ for $i = 1,\ldots,d-2$. Additionally, the five bases provide another set of submatrices ${\cal M}'$ to reconstruct $\rho$. This set of submatrices results from also measuring the elements $\rho_{d-1,0}$ and $\rho_{0,d-1}$, which were not used in the construction of ${\cal M}$. The failure set for ${\cal M}'$ is the same as the failure set of ${\cal M}$ but since ${\cal M'} \neq {\cal M}$ we gain additional robustness. When we consider both sets of submatrices the total failure set is $\rho_{i,i} = 0$ and $\rho_{j,j} = 0$ for $i = 1,\ldots,d-2$ and $i \neq j \pm 1$. This is the exact same set found by Goyeneche~{\em et al.}~\cite{Goyeneche14}.

We generalize these ideas to measure a rank-$r$ state by designing $4r +1$ orthonormal bases that correspond to a rank-$r$ strictly-complete POVM. The algorithm for constructing these bases, for dimensions that are powers of two, is given in Algorithm~\ref{alg:rankr_GMB} in the Appendix. Technically, the algorithm produces unique bases for $r \leq d/2$ but, as mentioned before, since $d+1$ mutually unbiased bases are informationally complete, for $r \geq d/4$ one may prefer to measure the latter. The corresponding measured elements are the first $r$ diagonals of the density matrix. 

Given the first $r$ diagonals of the density matrix, we can reconstruct a state $\rho \in {\cal S}_r$ with a similar procedure as the one outlined for the five bases. First, choose the leading $\by{(r+2)}$ principle submatrix, $M_0$. The unmeasured elements in this submatrix are $\rho_{0,r+1}$ and $\rho_{r+1,0}$. By applying a unitary transformation we can bring $M_0$ into canonical form, and by using the rank condition from Eq.~\eqref{Schur_rank} we can solve for the unmeasured elements. We can repeat the procedure with the set of $\by{(r+2)}$ principle submatrices $M_i \in {\cal M}$ for for $i = 0,\ldots,d-r-1$ and 
\begin{equation}
M_{i} = \begin{pmatrix}
\rho_{i,i} & \cdots & \bm{\rho_{i,i+r+1}} \\
\vdots & \ddots & \vdots  \\
\bm{\rho_{i+r+1,i}} & \cdots &  \rho_{i+r+1,i+r+1}
\end{pmatrix}.
\end{equation}
From $M_i$ we can reconstruct the elements $\rho_{i,i+r+1}$, which form the $(r+1)$st diagonal. We then repeat this procedure choosing the set of $\by{(r+3)}$ principle submatrices to reconstruct the $(r+2)$nd diagonal and so on until all diagonals have been reconstructed. This shows the measurements are rank-$r$ complete and by Proposition~1, since we also measure the computational bases, the POVM is also rank-$r$ strictly-complete. 

The failure set corresponds to the set of states with singular $\by{r}$ principal submatrix
\begin{equation}
A_i = \begin{pmatrix}
\rho_{i+1,i+1} & \cdots & \rho_{i,i+r} \\
\vdots & \ddots & \vdots \\
\rho_{i+r,i} & \cdots & \rho_{i+r,i+r}
\end{pmatrix},
\end{equation}
for $i = 1,\ldots,d-r-1$. This procedure also has robustness to this set since, as in the case of $r = 1$, there is an additional construction ${\cal M}'$. The total failure set is then when $A_i$ is singular for $i = 0,\ldots,d-r-1$ and $A_j$ is singular for $j \neq i \pm 1$.

If we are given a state that is close to pure, similarly to the POVM of Example~1, we can iteratively create estimates of the state. The first five bases form a rank-$1$ strictly-complete measurement which gives most of  the  information about the state. Measuring the next four bases forms a rank-$2$ strictly-complete which gives a first-order estimate of the state. We can repeat, measuring more series of four bases, to get higher-order estimates of the state. 

\section{Summary and Conclusions}\label{sec:conclusions}
We have studied the problem of QST under the assumption that the state is pure and more generally has rank $\leq r$, for a permissible $r$. With this prior information, we can design special measurements that are more efficient than measurements for a general quantum state, that is, POVMs with less than $d^2$ elements. There are two notions of completeness for these measurements: rank-$r$ complete and rank-$r$ strictly-complete. Rank-$r$ complete measurements uniquely identify states within the set of rank $\leq r$ states while rank-$r$ strictly-complete measurements uniquely identify states within the set of all physical states, of any rank. The latter type of measurements are only possible due to the positivity property of the density matrix. Strict-completeness has a practical implication for estimating the state of system in the presence of noise.  Although the set of rank-$r$ states is nonconvex, using rank-$r$ strictly-complete POVMs we are able to use convex optimization programs to estimate the state of the system. Such programs ensure robust estimation of the quantum state. 

Generally, it is difficult to asses if an arbitrary POVM satisfies one of these completeness relations. In this work we studied POVMs that allow for the reconstruction of a few density matrix elements, EP-POVMs. In this situation the problem of QST is reduced to density matrix completion. We developed tools to determine if a given EP-POVM is rank-$r$ complete or rank-$r$ strictly-complete based on properties of the Schur complement. These tools provide a unified framework for all EP-POVMs. We applied them to previous constructions of POVMs by Flammia~{\em et al.}~\cite{Flammia05} and by Goyeneche~{\em et al.}~\cite{Goyeneche14}, and showed that they are rank-1 strictly-complete POVM. The framework  also allow us to construct new rank-$r$ strictly-complete EP-POVMs and we constructed two examples.  

One could use rank-$r$ strictly-complete POVMs, such as the ones presented in Sec.~\ref{sec:examples}, to iteratively probe highly-pure quantum states. A rank-$1$ strictly-complete POVM could be used to produce an estimate of the dominant eigenvalue, as was shown by Goyeneche~{\em et al.}~\cite{Goyeneche14}. We can use our generalization for rank-$r$ strictly-complete POVMs to produce more accurate estimates, when needed. For example, a rank-$2$ strictly-complete POVM, such as the ones introduced in Sec.~\ref{sec:examples}, would produce an estimate corresponding to the state's two largest eigenvalues. We could continue to produce more accurate estimates but at some point the eigenvalues will be so small that other sources of noise will dominate. In future work we plan to explore how one can use such an iterative procedure to certify the number of dominant eigenvalues in the state without performing full quantum tomography. 

This work provides a full understanding of element-probing measurements but these are only a subset of all possible measurements. There are non EP-POVMs that are rank-$r$ complete and rank-$r$ strictly-complete but our framework cannot be applied to them. This general case will be studied in subsequent work as well as the relation of this approach to compressed sensing techniques~\cite{Kalev15-2}. 

We focused here exclusively on QST and implicitly assumed that the POVMs are implemented perfectly. However, any experimental implementation of POVMs would be imperfect and thus the tomographic measurement record will be polluted by noise due to such imperfections. This motivates the necessity of self-consistent tomography procedure that includes quantum-process tomography and quantum-detector tomography. These important issues will be studied elsewhere.

\begin{acknowledgments}
This work was supported by NSF Grants PHY-1212445, PHY-1521016, and PHY-1521431.
\end{acknowledgments}

\begin{widetext}
\appendix
\section{}
\begin{algorithm}[H]
\caption{Construction of $4r +1$ bases that compose a rank-$r$ strictly-complete POVM}
 \label{alg:rankr_GMB}
\begin{enumerate}\label{alg:rankr_GMB}

\item{Construction of the first basis:}
\item[]{The choice of the first basis is arbitrary, we denote it by $\mathbbm{B}_0 = \{ |0 \rangle, | 1 \rangle, \ldots , | d-1 \rangle \}$. This basis defines the representation of the density matrix. Measuring this basis corresponds to the measurement of the all the elements on the 0th diagonal of $\rho$.}

\item{Construction of the other $4r$ orthonormal bases:}
\item[]{{\bf for} $k \in [1, r]$, {\bf do}}
\begin{itemize}
\item[]{Label the elements in the $k$th diagonal of the density matrix by $\rho_{m,n}$ where $m = 0,\ldots,d - 1 - k$ and $n = m + k$.}
\item[]{For each element on the $k$th and $({d-k})$th diagonal, $\rho_{m,n}$, associate two, two-dimensional, orthonormal bases,
\begin{align} \label{2dim_bases}
\mathbbm{b}^{(m,n)}_{x}=&\Bigl\{| x_{m,n}^{\pm} \rangle = \frac{1}{\sqrt{2}} \left( | m \rangle \pm | n\rangle \right)\Bigr\}, \nonumber \\
\mathbbm{b}^{(m,n)}_{y}=&\Bigl\{| y_{m,n}^{\pm} \rangle = \frac{1}{\sqrt{2}} \left( | m \rangle \pm \ii | n \rangle \right)\Bigr\},
\end{align}
for allowed values of $m$ and $n$.}

\item[]{Arrange the matrix elements of the $k$th diagonal and $({d-k})$th diagonal into a vector with $d$ elements 
\begin{equation}
\vec{v}(k) = ( \underbrace{\rho_{0,k}, \ldots, \rho_{d-1-k,d-1}}_\text{$k$th diagonal elements},\underbrace{\rho_{0,d-i},\ldots,  \rho_{k-1,d-1}}_\text{$({d-k})$th diagonal elements})\equiv(v_1(k),\ldots,v_d(k)). 
\end{equation}}
\item[]{Find the largest integer $Z$ such that $\frac{k}{2^{Z}}$ is an integer.}
\item[]{Group the elements of $\vec{v}(k)$ into two vectors, each with $d/2$ elements, by selecting $\ell = 2^Z$ elements out of $\vec{v}(k)$ in an alternative fashion,
\begin{align}
\vec{v}^{(1)}(k)  &= ( v_1, \ldots, v_\ell, v_{2\ell+1}, \ldots, v_{3\ell}, \ldots, v_{d-2\ell+1}, \ldots, v_{d-\ell} ) =(\rho_{0,i},\ldots,\rho_{\ell,i+\ell},\ldots),\nonumber \\
\vec{v}^{(2)}(k) &=( v_{\ell+1}, \ldots, v_{2\ell}, v_{3\ell+1}, \ldots, v_{4\ell}, \ldots, v_{d-\ell+1}, \ldots, v_{d})=(\rho_{\ell+1,i+\ell+1},\ldots, \rho_{2\ell,i+2\ell},\ldots)\nonumber
\end{align}}

\item[]{{\bf for} $j=1,2$ {\bf do}}
\begin{itemize}
\item[]{Each element of $\vec{v}^{(j)}(k)$ has two corresponding bases $\mathbbm{b}^{(m,n)}_{x}$ and $\mathbbm{b}^{(m,n)}_{y}$ from Eq.~\eqref{2dim_bases}.}
\item[]{Union all the two-dimensional orthonormal $x$-type bases into one basis
\begin{equation}
\mathbbm{B}^{(k;j)}_{x}=\bigcup_{\rho_{m,n}\in\vec{v}^{(j)}(k)} \mathbbm{b}^{(m,n)}_{x}.
\end{equation}
Union all the two-dimensional orthonormal $y$-type bases into one basis
\begin{equation}
\mathbbm{B}^{(k;j)}_{y}=\bigcup_{\rho_{m,n}\in\vec{v}^{(j)}(k)} \mathbbm{b}^{(m,n)}_{y}.
\end{equation}
The two bases $\mathbbm{B}^{(k;j)}_{x}$ and $\mathbbm{B}^{(k;j)}_{y}$ are orthonormal bases for the $d$-dimensional Hilbert space. }
\end{itemize}
\item[]{{\bf end for}}
\item[]{By measuring $\mathbbm{B}^{(k;j)}_{x}$ and $\mathbbm{B}^{(k;j)}_{y}$ for $j=1,2$ (four bases in total), we measure all the elements on the $k$th and $(d-k)$th off-diagonals of the density matrix.}
\end{itemize}
\item[]{{\bf end for}}
\end{enumerate}
\end{algorithm}

\end{widetext}


\begin{thebibliography}{99}
\bibitem{footnote_tr}%
More generally, a POVM may give more information than certain matrix elements. This information could, and sometime should, also be taken into account for the reconstruction of the density matrix. This case is beyond the scope of the current paper.
\bibitem{footnote_rank}%
Rank-$r$ complete POVMs can also completely specify states whose rank is smaller than $r$. If the state of the system has rank smaller than $r$,  then any $\by{r}$ principal submatrix of $\rho$ is singular. In this case, we chose a the largest principal matrix, $A$, which is nonsingular. Its dimension corresponds to the rank of the density matrix.
\end{thebibliography}
\end{document}